\documentclass[conference]{IEEEtran}
\IEEEoverridecommandlockouts
\usepackage{amsmath,amssymb,amsfonts}
\usepackage{xcolor}
\usepackage{graphicx}
\usepackage{acronym}
\usepackage[T1]{fontenc}
\usepackage[english]{babel}
\usepackage[normalem]{ulem}
\usepackage{algorithm}
\usepackage[noend]{algpseudocode}
\usepackage{footnote}
\usepackage{tikz}
\usepackage{subfigure}
\usepackage[utf8x]{inputenc}
\usepackage{textcomp}
\usepackage{lettrine}
\usepackage{txfonts}
\usepackage{siunitx}
\usepackage{makecell}
\usepackage{fancyhdr,setspace,ifthen}
\usepackage[verbose]{cite}
\usepackage{graphicx,psfrag,epsfig}
\usepackage{amssymb}
\usepackage{array}
\usepackage{hhline}
\usepackage{epic}
\usepackage{eepic}
\usepackage{enumerate}
\usepackage{color}
\usepackage{comment}
\usepackage{subfig} 
\usepackage{url}
\usepackage{indentfirst}
\usepackage{float}
\usepackage{booktabs}
\usepackage{lscape}
\usepackage{caption}
\usepackage{longtable}
\usepackage{paralist}
\usepackage[shortlabels]{enumitem}
\usepackage{pifont}

\definecolor{Yellow}{rgb}{1,0.9,0.7}
\definecolor{Pink}{rgb}{1,0.85,0.85}
\definecolor{AntiqueWhite}{rgb}{0.9,0.9,0.9}

\def\BibTeX{{\rm B\kern-.05em{\sc i\kern-.025em b}\kern-.08em
    T\kern-.1667em\lower.7ex\hbox{E}\kern-.125emX}}
\begin{document}

\title{Adaptive Workload Distribution for Accuracy-aware DNN Inference on Collaborative Edge Platforms
\thanks{\textbf{This manuscript is accepted to be published in 29th Asia and South Pacific Design Automation Conference (IEEE ASP-DAC 2024).\\
\indent The authors gratefully acknowledge funding from EU Horizon 2020 Research and Innovation Programme under the Marie Sk\l{}odowska Curie grant No. $956090$ (APROPOS), Nokia Foundation and Kaute Saatio.}}
}

\author{\IEEEauthorblockN{Zain Taufique}
\IEEEauthorblockA{\textit{University of Turku}\\
zatauf@utu.fi}
\and
\IEEEauthorblockN{Antonio Miele}
\IEEEauthorblockA{\textit{Politecnico di Milano}\\
antonio.miele@polimi.it}
\and
\IEEEauthorblockN{Pasi Liljeberg}
\IEEEauthorblockA{\textit{University of Turku}\\
pasi.liljeberg@utu.fi}
\and
\IEEEauthorblockN{Anil Kanduri}
\IEEEauthorblockA{\textit{University of Turku}\\
spakan@utu.fi}
}

\maketitle

\acrodef{api}[API]{Application Programming Interface}
\acrodef{cpu}[CPU]{Central Processing Unit}
\acrodef{dvfs}[DVFS]{Dynamic Voltage/Frequency Scaling}
\acrodef{hmp}[HMP]{Heterogeneous Multi-Processing}
\acrodef{hsa}[HSA]{Heterogeneous System Architecture}
\acrodef{os}[OS]{Operating System}
\acrodef{tdp}[TDP]{Thermal Design Power}
\acrodef{gn}[GN]{Gateway Node}
\acrodef{ln}[LN]{Local Node}
\acrodef{iot}[IoT]{Internet of Things}
\acrodef{ml}[ML]{Machine Learning}
\acrodef{qos}[QoS]{Quality of Service}
\acrodef{ftp}[FTP]{Fused Tile Partitioning}
\acrodef{aofl}[AOFL]{Adaptive Optimal Fused-layer}
\acrodef{socs}[SoCs] {System-on-Chips}
\acrodef{rtm}[RTM]{Run-time Resource Management}
\acrodef{dnn}[DNN]{Deep Neural Network}
\acrodef{dop}[DoP]{Degree of Parallelism}
\acrodef{ppw}[PPW]{Performance per Watt}
\acrodef{dnn}[DNN]{Deep Neural Network}
\acrodef{os}[OS]{Operating System}
\acrodef{fsm}[FSM]{Finite State Machine}
\acrodef{lan}[LAN]{Local Area Network}
\acrodef{wlan}[WLAN]{Wireless Local Area Network}
\acrodef{elan}[ELAN]{Ethernet Local Area Network}
\acrodef{ilp}[ILP]{Integer Linear Programming}
\acrodef{eeg}[EEG]{Electroenceophelogram} 

\acused{cpu}

\begin{tikzpicture}[remember picture, overlay]
  \node[rotate=90, right, text width=20cm, align=center, anchor=south] at (current page.east) {Accepted to be published in 29th Asia and South Pacific Design Automation Conference (IEEE ASP-DAC 2024)};
\end{tikzpicture}

\begin{abstract}
DNN inference can be accelerated by distributing the workload among a cluster of collaborative edge nodes. Heterogeneity among edge devices and accuracy-performance trade-offs of DNN models present a complex exploration space while catering to the inference performance requirements. In this work, we propose adaptive workload distribution for DNN inference, jointly considering node-level heterogeneity of edge devices, and application-specific accuracy and performance requirements. Our proposed approach combinatorially optimizes heterogeneity-aware workload partitioning and dynamic accuracy configuration of DNN models to ensure performance and accuracy guarantees. We tested our approach on an edge cluster of Odroid XU4, Raspberry Pi4, and Jetson Nano boards and achieved an average gain of 41.52\% in performance and 5.2\% in output accuracy as compared to state-of-the-art workload distribution strategies. 

\end{abstract}
\noindent


\section{Introduction}
\begin{figure}[t]
\centering
\vspace{-3pt}
\includegraphics[width=0.95\columnwidth]{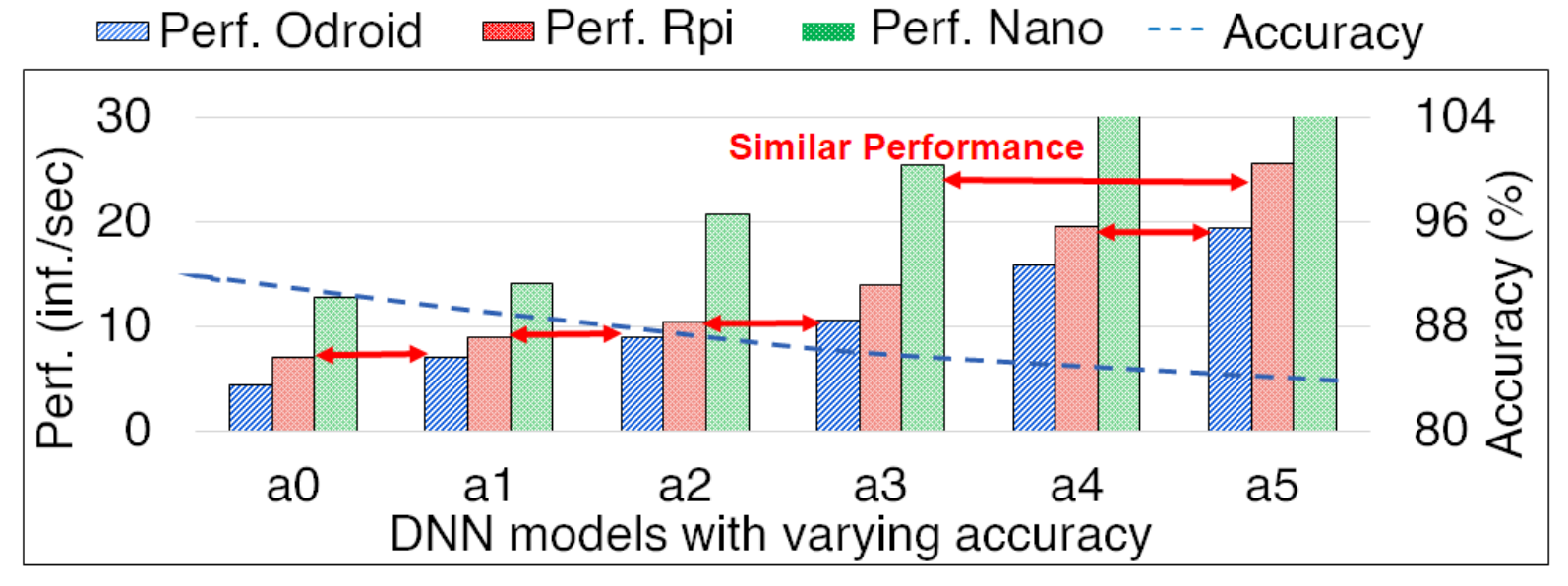}
\vspace{-6pt}
\caption{Accuracy-perfomance trade-offs of inferring MobileNetV2 across different edge devices.}
\vspace{-21pt}
\label{fig.intro}
\end{figure}

\acp{dnn} provide smart and autonomous services in application domains such as \ac{iot}, by processing continuous input data streams from multiple sources. 
The compute-intensive \ac{dnn} inference requests are typically offloaded to resourceful cloud servers to handle the compute requirements. However, continuous data transfer between user-end edge devices and remote cloud servers affects quality-of-service with increased inference latency, disconnection risks, and data privacy concerns \cite{AutoDice}. To address these challenges, \ac{dnn} inference requests can be distributed across a cluster of locally connected edge devices for collaborative execution \cite{Video_data_Parallel}. Edge clusters are increasingly becoming heterogeneous, where the connected devices have different hardware architectures and computational capabilities \cite{Legion, AutoScale}. On the other hand, widely used pruned versions of \ac{dnn} models present different accuracy-performance trade-offs \cite{Even_Apx}. Combining the node-level heterogeneity of edge devices and accuracy-performance trade-offs of different \ac{dnn} models exposes a significantly wider Pareto space to navigate through for efficient workload distribution. 
We present the performance diversity among a heterogeneous cluster of edge devices viz., Odroid XU4, Raspberry Pi4, and Jetson Nano boards, while running image classification on pre-trained MobileNetV2 \cite{MobileNetV2}. We executed 6 different versions of MobileNet models (a0-a5), exhibiting varying levels of accuracy (92.5\% - 82.9\%) with pruned model parameters by reducing the multiplication width \cite{tflite}. Figure \ref{fig.intro} shows the performance and accuracy of different MobileNet models (a0-a5) across heterogeneous edge devices. While each edge device has specific inherent compute capabilities, different devices can still deliver a similar performance (inferences-per-second) by tweaking the accuracy of the \ac{dnn} model. For example, Jetson Nano performs better than Odroid XU4 and Raspberry Pi4 across different models. However, both Raspberry Pi4 and Odroid XU4 can provide a similar performance to that of Jetson Nano (shown in red arrows in Figure \ref{fig.intro}), although at a lower accuracy. This demonstrates a non-intuitive accuracy-performance-heterogeneity optimal space that can be exploited for efficient workload distribution. 

Existing strategies for distributing \ac{dnn} inference workloads across edge clusters use input data partitioning \cite{Video_data_Parallel, encryption} and/or \ac{dnn} model splitting \cite{EdgeLD, MoDNN, DeepThings, Legion}, depending on application-specific scenarios. However, these strategies consider (i) accuracy-performance trade-offs assuming homogeneous edge nodes, or (ii) heterogeneous edge nodes ignoring accuracy-performance trade-offs.
Consequently, these approaches have limited efficacy in handling dynamic scenarios such as run-time workload variation and device availability. Addressing these challenges necessitates an intelligent workload distribution policy that combinatorially optimizes workload partitioning and accuracy configuration-- to meet performance and accuracy constraints, considering heterogeneity and availability of edge nodes. 




In this work, we present an adaptive workload distribution strategy to dynamically partition, distribute, and set the accuracy levels of \ac{dnn} inference workloads executed over collaborative heterogeneous edge clusters. 
We design an edge cluster framework that supports workload distribution among connected heterogeneous edge nodes by monitoring run-time system and workload dynamics. Our workload distribution policy makes online decisions on workload partitioning, distribution, and accuracy configuration of \ac{dnn} inference requests, optimizing the collaborative resource utilization across the edge cluster, while meeting the performance and accuracy requirements. Our proposed policy sets the accuracy levels of \ac{dnn} models through a dynamic selection of appropriate \ac{dnn} kernels from a pool of pre-trained models that exhibit different performance-accuracy trade-offs \cite{Even_Apx}. Our contributions are: \begin{inparaenum} [(i)] 
    \item design of edge cluster framework supporting run-time monitoring of system dynamics and workload distribution,
    \item heterogeneity aware adaptive workload distribution policy exploiting accuracy-performance trade-offs
    \item deployment of our proposed framework on real hardware edge cluster (Odroid XU4, Raspberry Pi4, and Jetson Nano) and evaluation against state-of-the-art workload distribution strategies.
\end{inparaenum}

\vspace{-1pt}
\section{Motivation and Related Work} \label{sec.Background}
\vspace{-1pt}

\begin{figure}[t]
\centering
\vspace{-8pt}
\includegraphics[width=0.8\columnwidth]{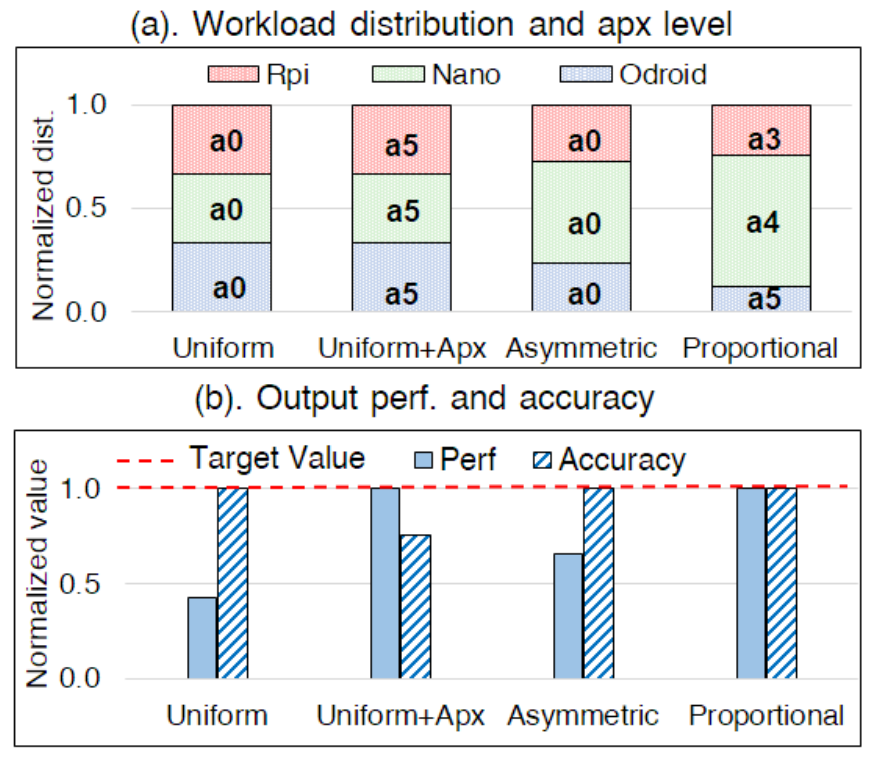}
\vspace{-3pt}
\caption{Workload distribution through different strategies. (a) workload distribution and selected approximation level on each board. (b). Overall output performance and accuracy.}
\vspace{-15pt}
\label{fig.mot2}
\vspace{-6pt}
\end{figure}

 \subsection{Motivation} \label{sec.Motivation}
Traditionally, \ac{dnn} workloads are distributed in an edge cluster: \begin{inparaenum}[(i)]
    \item \textit{uniformly} -- equally distributing the workload across different devices\cite{MoDNN, DeepThings},
    \item \textit{asymmetrically} -- distributing the workload proportional to the devices' compute capabilities \cite{EdgeLD, APED, Legion}, or
    \item \textit{uniformly with approximation} -- evenly distributing the workload and aggressively approximating to meet performance requirements \cite{Even_Apx}\end{inparaenum}.
Distributing the workload uniformly in heterogeneous environments causes an unbalanced use of resources that leads to performance violations in harsh workload scenarios. Asymmetrically distributing the workload according to node capability provides relatively better performance by optimizing the usage of the available resources. However, the lack of approximation limits this strategy's performance to the rated capacity of the available nodes. Finally, the uniform distribution with approximation can achieve the required performance of intense workloads but aggressively sacrificing the accuracy may violate the accuracy requirements of the workload. Given the limitations of the aforementioned approaches, a \textit{Proportional} workload distribution strategy is required to determine optimal partitioning that meets the performance requirements within the minimum possible approximation level. We demonstrate the distribution of a compute-intensive \ac{dnn} inference workload using the \textit{uniform, uniform+apx, asymmetric} and \textit{proportional} approaches in an edge cluster of Raspberry Pi4, Jetson Nano, and Odroid XU4 boards as shown in Figure \ref{fig.mot2} (a). 
For each strategy, we show the stacked percentile of workload distributed along with the selected \ac{dnn} model across different devices. The block size within the stack shows the workload partition size, the color of the block represents the device, and the model number (a0-a5) represents the approximation level of the selected model.
We present the output accuracy and performance of all these strategies in Figure \ref{fig.mot2} (b). The \textit{Uniform} and \textit{Uniform+apx} strategies distribute the workload equally among the available boards. The \textit{Uniform} approach cannot achieve the performance but is successfully meeting the accuracy requirements because it does not approximate the workload. The \textit{Uniform+apx} is utilizing approximation to meet the performance requirements, but its aggressive use of approximation violates the accuracy requirements of the workload. The workload is distributed unevenly in both \textit{Asymmetric} and \textit{Proportional} approaches. However, the \textit{Asymmetric} approach cannot meet the required performance because the workload requirements are far higher than the limitation of the available devices. The \textit{Proportional} strategy is heterogeneity-aware and achieves the target accuracy and performance by intelligently partitioning and approximating the workload proportionally to each board's computational capacity. With the insights from the demonstrative example in Figure \ref{fig.mot2}, we posit that 
\begin{inparaenum} [(i)]
    \item workload partitioning and accuracy configuration should be fine-tuned together to meet performance and accuracy requirements, and
    \item edge node-level heterogeneity-awareness is the key for jointly actuating workload partitioning and approximation.
\end{inparaenum} Our goal is to design an adaptive workload distribution policy that outlines the \textit{Proportional} strategy to ensure optimal usage of the edge cluster while meeting the \ac{dnn} application requirements.

\begin{figure*}[t]
\centering
\vspace{-3pt}
\includegraphics[width=0.8\textwidth]{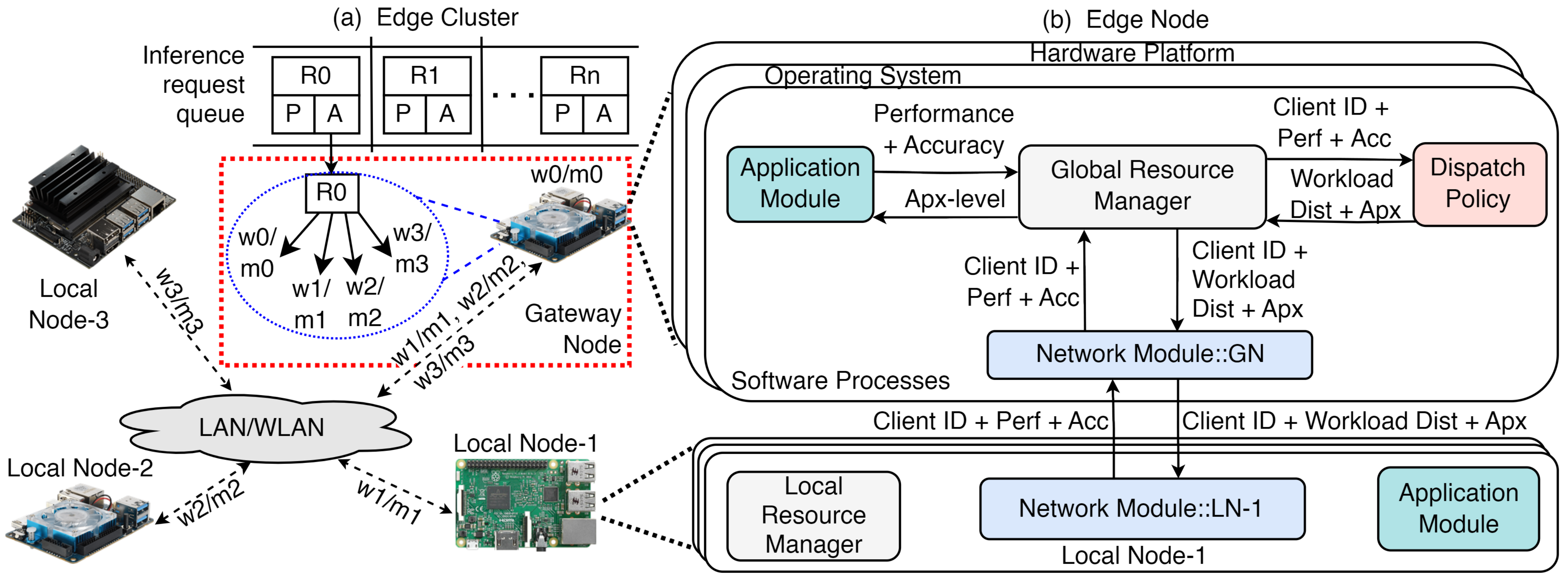}
\vspace{-3pt}
\caption{System Diagram showing (a) edge cluster (b) the node-level software and hardware modules}
\vspace{-15pt}
\label{fig.system_diagram}
\end{figure*}

\vspace{-3pt}
\begin{table}[b]
  \caption{Comparison of our and existing approaches.}
  \label{tab.comparison}
  \centering
  \setlength\tabcolsep{4.85pt}
  \scalebox{0.85}{
    \begin{tabular}{|l|c|c|c|c|c|c|c|c|c|}
      \hline
      & \cite{AutoDice} & \cite{EdgeLD} & \cite{MoDNN} & \cite{DeepThings} & \cite{Legion} & \cite{Even_Apx} & \cite{APED} & \cite{DMP} & \textbf{Our} \\
      \hline
      Perf. aware & \ding{51} & \ding{51} & \ding{51} & \ding{51} & \ding{51} & \ding{51} & \ding{51} & \ding{51} & \ding{51} \\
      \hline
      Acc. aware & \ding{51} & \ding{56} & \ding{56} & \ding{51} & \ding{51} & \ding{51} & \ding{56} & \ding{56} & \ding{51} \\
      \hline
      Heter. aware & \ding{51} & \ding{51} & \ding{56} & \ding{56} & \ding{51} & \ding{56} & \ding{51} & \ding{51} & \ding{51} \\
      \hline
      Adaptive & \ding{51} & \ding{56} & \ding{56} & \ding{51} & \ding{51} & \ding{51} & \ding{51} & \ding{51} & \ding{51} \\
      \hline
      Run-time & \ding{56} & \ding{56} & \ding{56} & \ding{56} & \ding{56} & \ding{51} & \ding{56} & \ding{56} & \ding{51} \\
      \hline
    \end{tabular}
  }
\end{table}

\subsection{Related Work} \label{sec.Related}
Run-time workload distribution of \ac{dnn} applications to achieve the target performance and accuracy while optimizing the usage of heterogeneous resources is a complex challenge. Most research works \cite{MoDNN, DeepThings} evenly partition the workload and are suitable for homogeneous edge clusters. In contrast, recent works \cite{EdgeLD, APED, Legion, AutoDice, DMP} consider heterogeneity for workload distribution with design-time knowledge of workload conditions and the available resources. Zhou et al. \cite{APED} and Legion \cite{Legion} presented adaptive frameworks that partition the workload based on the device performance and the network bandwidth but suffer if a device shuts down during the execution unexpectedly. AutoDice \cite{AutoDice} extended these solutions to form an autonomous framework that performs scenario-based workload partitioning with performance awareness, given that the user defines the available resources at design time, and these resources remain available during the entire execution cycle. Sina et al. \cite{Even_Apx} considered the availability of the resources and presented an accuracy-aware workload distribution strategy for the edge-cloud paradigm, but the solution considers resource homogeneity while distributing the workload. We propose an advanced workload distribution framework that exceeds the state-of-the-art approaches by having performance, accuracy, and heterogeneity awareness and adapting to varying application and resource scenarios, as shown in Table \ref{tab.comparison}.

\section{Adaptive Workload Distribution} \label{sec.methodology}


\subsection{Framework Overview}
We have designed an edge cluster framework comprising heterogeneous hardware nodes connected with wLAN, as shown in Figure \ref{fig.system_diagram} (a). Each node within the cluster handles incoming applications, network connection, workload distribution, and resource allocation through \ac{os} level support. Figure \ref{fig.system_diagram} (b) presents an overview of the designed software infrastructure in an edge node. We have considered a hierarchical architecture for the edge cluster, including a leader \ac{gn} and multiple followers \ac{ln}. The \ac{gn} receives a queue of inference requests ($R_{0}$,$R_{1}$,...,$R_{n}$) where each request has a batch of images with specific performance and accuracy requirements (\textit{P|A}). We assume that the inference requests can be serviced by selecting a model from a set of models exhibiting variable accuracy-performance trade-offs. The \ac{gn} partitions the workload in a data-parallel manner ($w_{0}$,$w_{1}$,...,$w_{n}$) and selects the appropriate model  ($m_{0}$,$m_{1}$,...,$m_{n}$), enabling the \ac{ln}s to meet the required performance and accuracy constraints. We have generalized our design to allow any node to be a \ac{gn} as per user requirements. The platform design for intelligent workload distribution is discussed in the following.

\noindent\textbf{Hardware Platform and \ac{os}.}
The distributed hardware platforms may comprise multiple heterogeneous multi-core nodes.
Each node supports run-time power monitoring through onboard power sensors or external monitoring equipment. 
We considered the edge node platforms with \ac{dvfs} actuation support to restrict the power consumption of the board below a maximum power consumption limit called \ac{tdp}, for thermal safety. Each edge node runs a Linux-based \ac{os} to enable interaction between the node hardware and software modules.
We use the \ac{os} interfaces at run-time to map the workload to the \ac{cpu} cores, monitor \ac{cpu} utilization and power consumption, and actuate \ac{dvfs} to maximize performance within fixed \ac{tdp}. The \textit{Application Module} loads the required libraries and kernels to perform the inference of the selected model within a given workload; and stores the resultant performance and accuracy measures. The hardware nodes require a network connection with ethernet or WiFi support to enable inter-board communication. The \textit{Network Module} enables the \ac{gn} to exchange data with a dedicated \ac{ln} at run-time and provides a list of available devices.
\newline \noindent\textbf{Workloads.} 
We have designed the framework to process \ac{dnn} inference workload with defined performance and accuracy requirements. We consider inference workload for streaming applications where the input data can be partitioned into smaller batches and distributed among the available nodes for parallel inference. We have considered scenario-based applications that have dynamic inference requests of variable input sizes, performance, and accuracy requirements. A common example of our considered applications is smart video surveillance \cite{Video_data_Parallel}, where the inference request data is a set of image frames that can be divided into small batches and distributed to the edge cluster for parallel inference. In our designed framework, each node has saved pre-trained models that can be used at run-time to provide different performances and accuracy. The \ac{gn} is designed to distribute the input data and select the optimal model for each node to meet the global accuracy and performance requirements of an inference request. As shown in Figure \ref{fig.system_diagram} (a), our framework saves an input requests queue in the \ac{gn} as a tuple vector, including inference request number R, required performance P, and required accuracy A. The \ac{gn} takes a request R1 from the queue, partitions it to four workloads ($w_1$,$w_2$,$w_3$,$w_4$) with an associated model number, and distributes it over the edge.

\begin{figure}[t]
\centering
\vspace{-12pt}
\includegraphics[width=0.99\columnwidth]{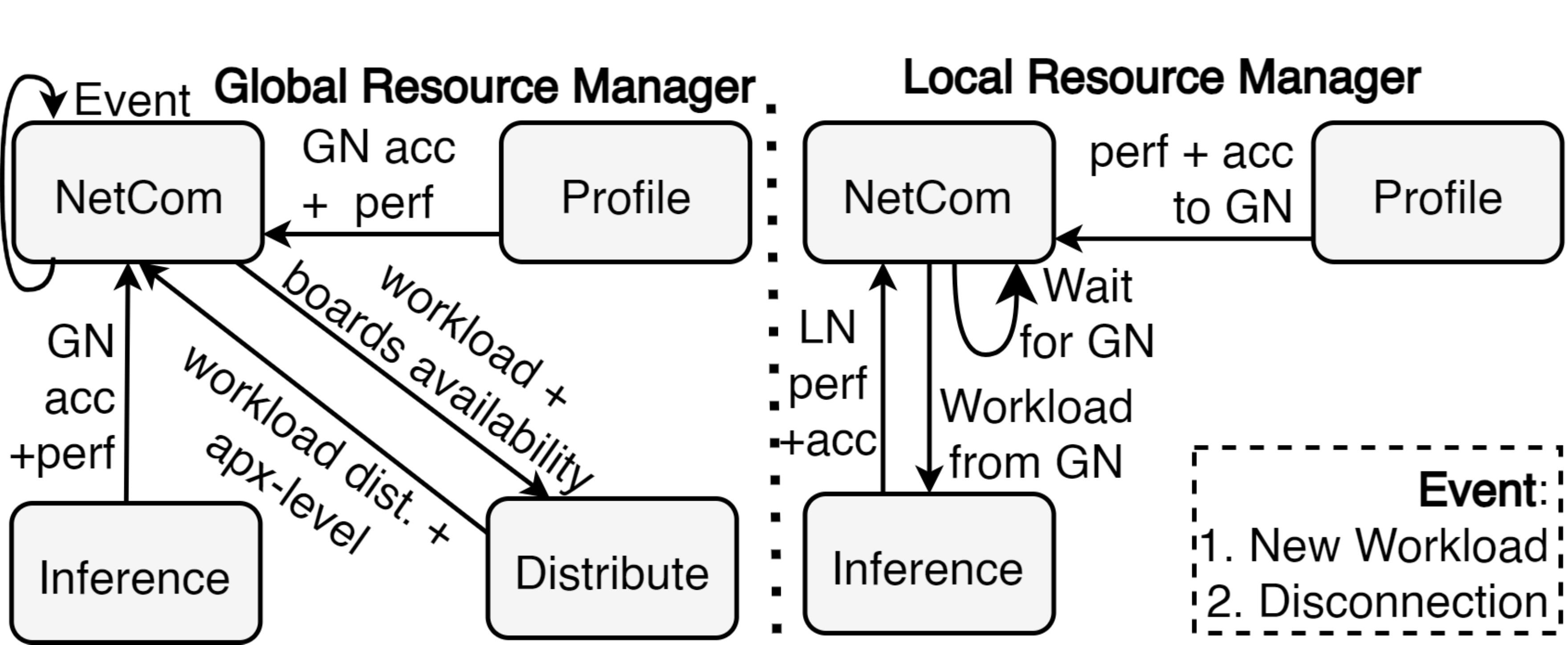}
\caption{State Machine of \ac{gn} and \ac{ln} Resource Manager.}
\vspace{-12pt}
\label{fig.fsm}
\end{figure}

\subsection{Resource Manager} We designed a distributed \textit{Resource Manager} comprising a global module on the \ac{gn} and a local module on each \ac{ln}. The behavior of both modules is represented as a \ac{fsm} in Figure~\ref{fig.fsm}.

\noindent\textbf{Global Resource Manager.} 
The \textit{Global Resource Manager} deployed in the \ac{gn} has four states, viz. \textit{Profile, NetCom, Distribute} and \textit{Inference}. In the \textit{Profile} state, we record the performance and accuracy of the \ac{gn} at different approximation levels with test data from the \textit{ImageNet} dataset to populate a profiling look-up table. After profiling, the \ac{fsm} transitions to \textit{NetCom} state to gather the connection status and profiling data of the connected \ac{ln}s using the \textit{Network Module}. The \ac{fsm} updates the profiling table with the received \ac{ln} data and waits for an event of either workload arrival or board disconnection. 
In case of a new workload, the \ac{fsm} receives the workload performance and accuracy requirements and transitions to the \textit{Distribute} state. The \textit{Distribute State} invokes the \textit{Dispatch Policy} (Section~\ref{sec.dispatch}) to receive the workload distribution with specified approximation levels and transitions to the \textit{NetCom} state. The \textit{NetCom} state broadcasts the workload distribution with the associated approximation levels to the relevant \ac{ln}s and transitions to the \textit{Inference} state. In the \textit{Inference} state, the \ac{fsm} triggers local inference of the received workload partition with associated approximation level and transitions to the \textit{NetCom} state. Finally, if a board disconnects at run-time, the \ac{fsm} again transitions to the \textit{Ditribute} state for new workload distribution and broadcasts the new distribution with updated approximation levels to all the available boards.


\noindent\textbf{Local Resource Manager.}
The \ac{fsm} starts with the \textit{Profile} state to record the performance and accuracy with test data and moves to the \textit{NetCom} state to send the profiled data to the \ac{gn}. The \ac{ln} waits till the workload and the required approximation level are received from the \ac{gn} and moves to the \textit{Inference} phase to execute the local inference. Finally, after the successful inference, the \ac{fsm} moves back to the \textit{NetCom} state to send the output accuracy and performance to the \ac{gn}.

\subsection{Dispatch Policy for Workload Distribution} \label{sec.dispatch}
The \textit{Dispatch Policy} is a \ac{gn} sub-process that takes insights from the profiling table recorded by the \textit{Resource Manager} to provide the run-time workload distribution and approximation.

\begin{figure}[t]
\centering
\vspace{-3pt}
\includegraphics[width=0.95\columnwidth]{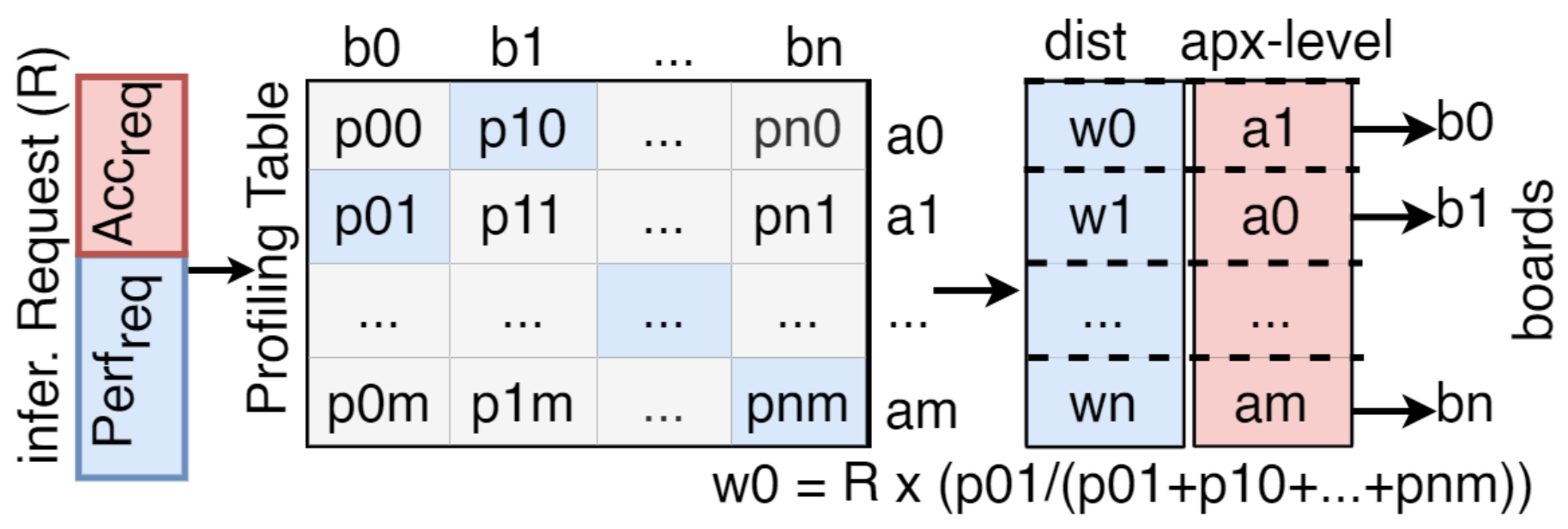}
\vspace{-3pt}
\caption{Workload distribution approach.}
\vspace{-18pt}
\label{fig.expTable}
\end{figure}

\noindent\textbf{Problem Formulation.}
The \textit{Global Resource Manager} records the performance values of all available boards $\texttt{b}_0, \texttt{b}_1, \ldots \texttt{b}_n$ in a profiling table $\texttt{profiling\_table}_{\texttt{mxn}}$ at different accuracy levels $a_0, a_1, \ldots, a_m$, where \texttt{n} is the number of boards and \texttt{m} is the lowest accuracy/highest approximation level. The columns of the profiling table represent the connected boards and the ascending row number shows the increasing approximation level as shown in Figure \ref{fig.expTable}. The \textit{Dispatch Policy} is invoked when a new inference request $R$ with requirements of performance $\texttt{Perf}_{\texttt{req}}$ and accuracy $\texttt{Acc}_{\texttt{req}}$ arrives in the system. The sum of recorded performances at each row represents the combined edge cluster performance at a given approximation level. The \textit{Dispatch Policy} calculates the percentage performance of each board against the cluster performance and proportionally splits the total workload performance requirement $\text{perf}_\text{req}$ as $\texttt{perf}_{\texttt{req}} = \sum_{i=1}^{n} \texttt{perf}\_b_{\texttt{req}}$ for each board. The \textit{Dispatch Policy} is expected to find the required performance and accuracy by selecting the performance values in the profiling table closest to the per-board performance requirements without aggressively approximating the workload and return the workload distribution as \(\text{w\_dist} = \text{w}_{0},\text{w}_{1}, \ldots, \text{w}_{n}\) with the associated approximation levels.


\noindent\textbf{Heuristics Model.}
The \textit{Dispatch Policy} prunes the exploration space to find the distribution that meets the workload requirements. 
Algorithm~\ref{alg:resource_manager_fsm} shows that the policy copies the data of the \texttt{profiling\_table} in a \texttt{pruned\_table} while ignoring the disconnected boards (Lines 3--5). The policy saves the cluster performance at each approximation level 
in a \texttt{perf\_vector} and stops when the recorded performance is equal to or higher than the required performance (Lines 6--9). It deletes all the remaining performance records at higher approximations to further reduce the exploration space (Lines 10--11). The policy calculates the board-level performance requirements and saves them in a vector \texttt{perf\_b$_{\texttt{req}}$[n]} (Lines 12--13). Then, it sends \texttt{perf\_b$_{\texttt{req}}$[]} and \texttt{pruned\_table[][]} to a dynamic programming algorithm DP$_{alg}$ to find the recorded performance values that are closest to the board-level performance requirements (Line 14).
In this context, we used a standard subset sum algorithm for an efficient recursive search with time complexity \texttt{O(n * m)}. 
The algorithm starts with the values of the highest approximation level and back-propagates row-by-row to reach the closest values. The algorithm returns the performance distribution vector \texttt{P\_dist} and the approximation vector \texttt{apx\_dist} for each board. Finally, the policy calculates workload distribution vector \texttt{W\_dist[]} by splitting the input data of the inference request $R$ proportional to the performance factor of each board (Lines 15--16).

\begin{algorithm}[t]
\vspace{-3pt}
\caption{Dispatch Policy}
\label{alg:resource_manager_fsm}
\small
\begin{algorithmic}[1]
\State {Input:} {$\text{profiling}\_\text{table}$, $\text{avail}\_\text{boards}$, \text{R}, $\text{Perf}_{\text{req}}$, $\text{Acc}_{\text{req}}$}
\Procedure{{Heuristics Model}}{}
    \For {\text{(m = 1 to max\_apx\_level)}}
        \For{\text{(n $\in$ \text{avail}\_\text{boards})}}
            \State {\text{pruned\_table[m][n] = profiling\_table[m][n]}}
        \EndFor
    \EndFor
    \For{\text{(index = 1 to m)}}
        \State {\text{total\_perf\_vector[index] = $\sum_{k=1}^{n}$ pruned\_table[index][k]}}
        \If{\text{perf\_vector[index] $ \ge \text{perf}_{\text{req}}$}}
            \State{\text{break;}}
        \EndIf
    \EndFor
    \For{\text{(m > index)}}
        \State{\text{pruned\_table[m][].delete()}}
    \EndFor

    \For{$(i \gets 0$ to $n)$}
      \State {$\text{perf}\_b_\text{req}[i] = \text{perf}_{\text{req}}$*pruned\_matrix[0][i]/perf\_vector[0]}
    \EndFor

    \State{\text{P\_dist[], apx\_dist[] = DP$_{alg}$(perf\_b$_\text{req}[]$,pruned\_table[m][n])}}
    \For{\text{l = 1 to n}}
        \State{\text{w\_dist[l] $\gets$ R * $\frac{\text{P\_dist[l]}}{\sum_{k=0}^{n}P\_dis[]}$}}
    \EndFor
\EndProcedure
\end{algorithmic}
\end{algorithm}

\section{Results and Evaluation} \label{sec.Results}
\subsection{Experimental setup}
We deployed our proposed framework on a cluster of four real hardware edge devices: Odroid XU4 (2x), Jetson Nano, and Raspberry Pi4. 

\noindent\textbf{Power Handling.} Following the technical datasheet of each board, we have set the \ac{tdp} limits of Odroid XU4, Raspberry Pi4, and Jetson Nano as 8W, 9W, and 10W. The Jetson Nano is equipped with in-board power sensors that we read at run-time for power monitoring. For Odroid XU4, we used \textit{Smart Power-3} analyzer \cite{SP3}, and for Raspberry Pi4, we used the external shunt resistor method as used in \cite{PiPower} to monitor power.

\noindent\textbf{Middleware Prototype.} 
Figure \ref{fig.system_diagram} shows the used setup where we have implemented the framework as a middleware in C++, and each board has a running Linux Ubuntu 20.04. We have implemented a POSIX-based client-server architecture connected via Ethernet, including multi-threaded server operations and static IP-based identification to enable data and command-sharing between the edge nodes. 
We implemented a generic macro-based source code for all boards with 10 source files, 7 header files, and about 2000 lines of source code.

\noindent\textbf{Workload Application.} 
We have considered image classification as the baseline \ac{dnn} application for our experimentation. The workload consists of a batch of images with labeled performance and accuracy requirements. The \ac{gn} partitions the number of images following the \textit{Dispatch Policy} (section \ref{sec.dispatch}) and distributes them among the available nodes. We use popular Tensorflow-Lite \cite{tflite} and OpenCV-Lite \cite{opencv_lite} libraries to load the input image and perform inference through pre-trained MobileNetV2 machine-learning models. 

\begin{figure}[t]
\centering
\vspace{-8pt}
\includegraphics[width=0.8\columnwidth]{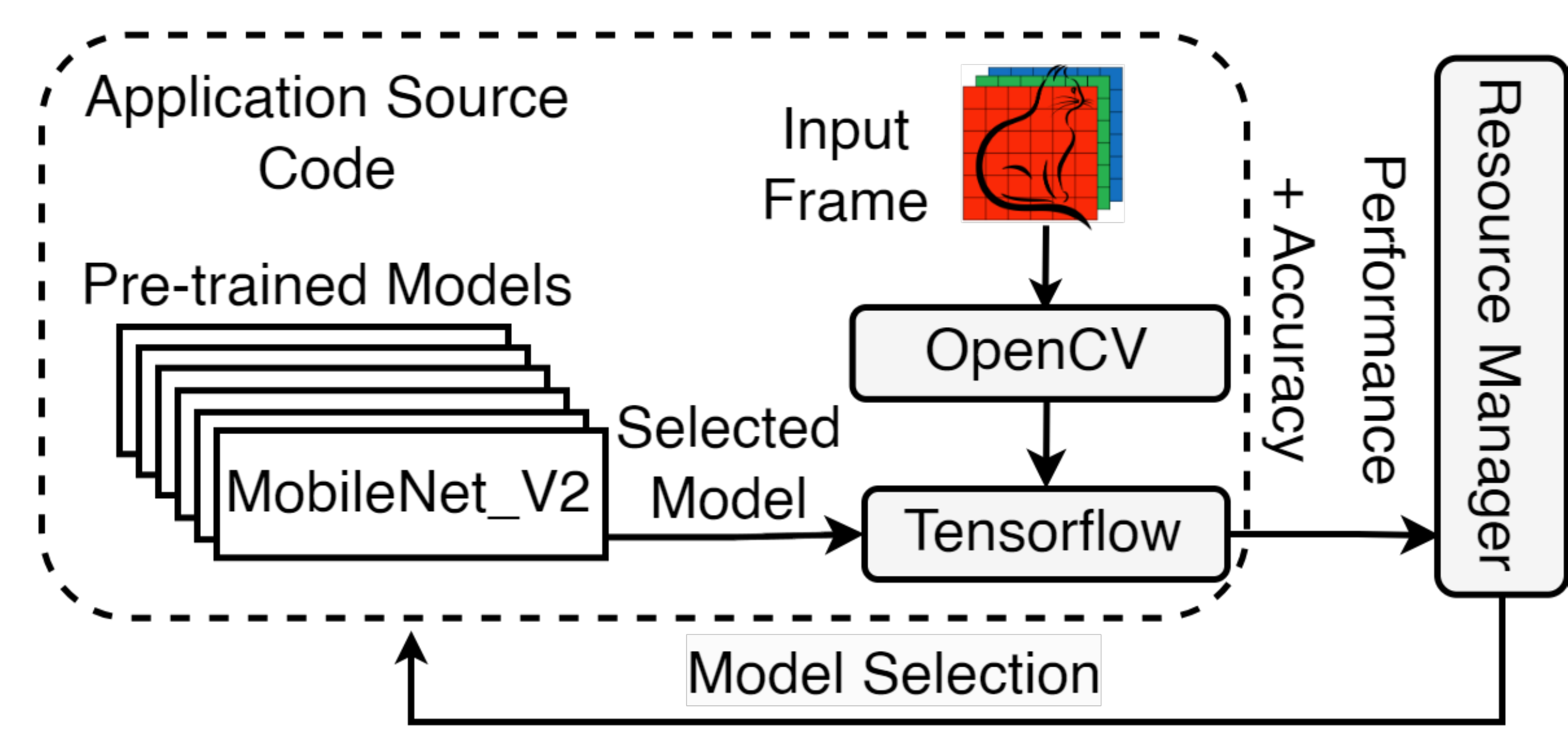}
\caption{Run-time approximation in the Application Layer}
\label{fig.approximation}
\vspace{-15pt}
\end{figure}

\noindent\textbf{Accuracy Configuration} We enforce accuracy configuration of \ac{dnn} models through a pre-trained model selection method similar to \cite{Even_Apx} and \cite{Hierarchical}. We used the six different MobileNetV2 models that are trained on the ImageNet dataset and are available online at the official Tensorflow-lite repository \cite{tflite}. These models are highly tunable and can achieve a range of accuracy and complexity trade-offs by adjusting the multiplier width parameter (alpha = 0.35, 0.5, 0.75, 1.0, 1.3, 1.4). Figure \ref{fig.approximation} presents the workflow of accuracy configuration through model selection. The \textit{Resource Manager} selects the model at run-time as determined by the \textit{Dispatch Policy}, and records the resultant accuracy and performance of the selected model.

\subsection{Experimental results}

\begin{figure*}[t]
\vspace{-3pt}
\centering
\includegraphics[width=0.9\textwidth]{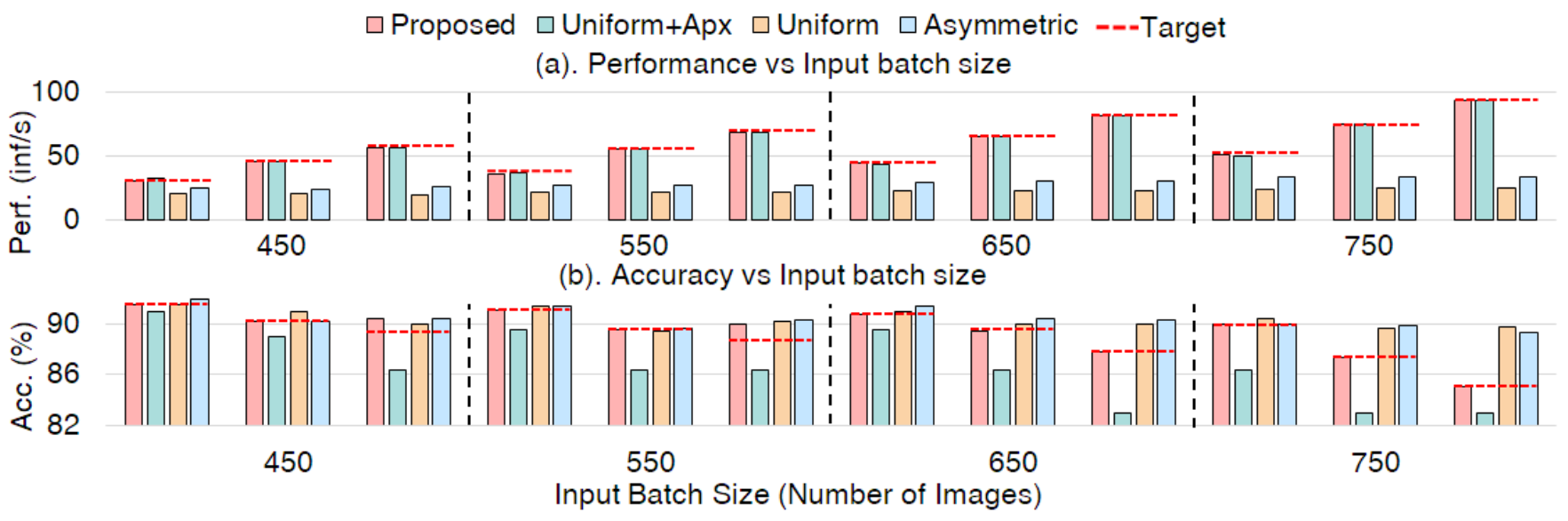}
\vspace{-6pt}
\caption{(a). Performance and (b). Accuracy (\%) with dynamic workload variation. Each combination has different performance and accuracy requirements, and input batch sizes.}
\label{fig.res_long}
\vspace{-12pt}
\end{figure*}


\noindent\textbf{Evaluation metrics}
We evaluate (i) performance as the number of inferences made per second and (ii) top-5\% output accuracy by comparing predicted labels to true labels and calculating the percentage of correctly classified samples. 

\noindent\textbf{Comparison baseline} 
We compared our results against the state-of-the-art workload distribution strategies including \cite{MoDNN} as \textit{Uniform} strategy for its even workload distribution in homogeneous edge clusters, \cite{Legion} as \textit{asymetric} approach for its non-uniform-workload distribution in heterogeneous edge cluster, and \cite{Even_Apx} as \textit{Uniform + Apx} distribution approach for doing uniform workload distribution while considering workload approximation.
We have profiled the performance of each device in terms of inference per second to formulate the profiling table. We have enhanced all of the compared strategies with \ac{dvfs} actuation to avoid the \ac{tdp} violation of each respective hardware board. We have only considered the workload distribution technique of these strategies and applied it to the input data partitioning for our implementation. 

\begin{figure}[t]
\centering
\vspace{-3pt}
\includegraphics[width=0.9\columnwidth]{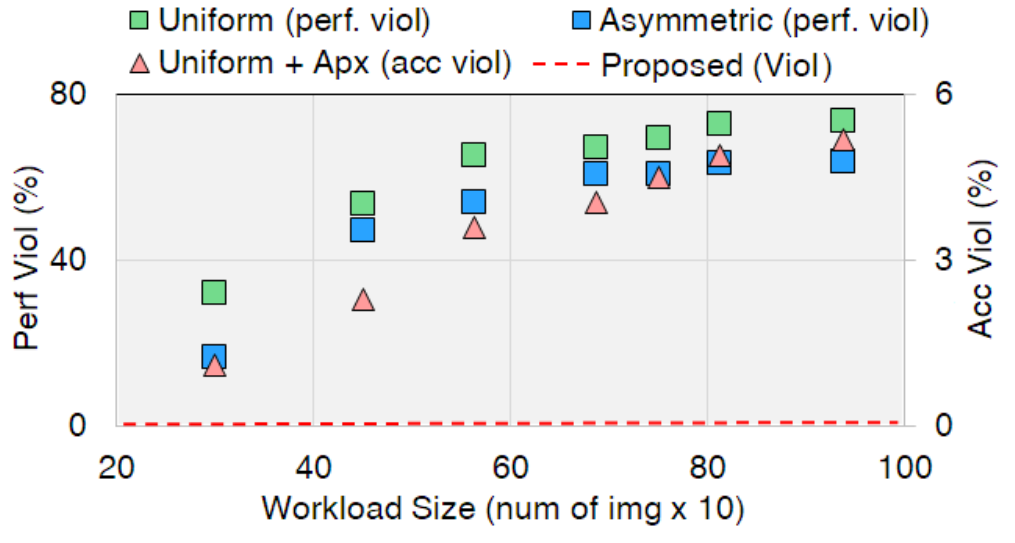}
\vspace{-9pt}
\caption{Performance and accuracy violations with varying input size (acc. viol. concern only uniform+apx and proposed).}
\label{fig.res_input_size}
\vspace{-15pt}
\end{figure}

\noindent\textbf{Varying Workload Scenarios.}
We compared each strategy under dynamic workload variations i.e. different workloads with variable performance and accuracy requirements, and input batch sizes. We present the output performance and accuracy of each experiment in Figure \ref{fig.res_long} (a) and (b), respectively. We experiment with four different input batch sizes, and three different performance and accuracy requirements for each batch. The results show that the \textit{Uniform + Apx} strategy meets the performance for each experiment while violating the minimum accuracy target due to aggressive approximation of the given workload. The \textit{Asymmetric} and \textit{Uniform} strategies generate more accurate results but are unable to meet the performance thresholds because the requirements are beyond the rated capabilities of the given edge cluster. 
Finally, the proposed strategy finds the optimal partitioning and approximation ensuring that the performance thresholds are met at the maximum output accuracy. Our approach yields an average performance gain of 52.63\% and an accuracy improvement of 3.9\% for the given set of experiments.
Figure \ref{fig.res_input_size} shows the average performance violations (\%) of \textit{Uniform} and \textit{Asymmetric} strategies and accuracy violations (\%) of \textit{Uniform + Apx} strategy under the aforementioned workload variation. We calculated the violation as the amount of time a strategy is unable to meet the target performance or accuracy in an execution cycle. The proposed strategy minimizes performance and accuracy violations on average by 41.52\% and 5.2\%.

\begin{figure}[t]
\centering
\vspace{-18pt}
\includegraphics[width=0.9\columnwidth]{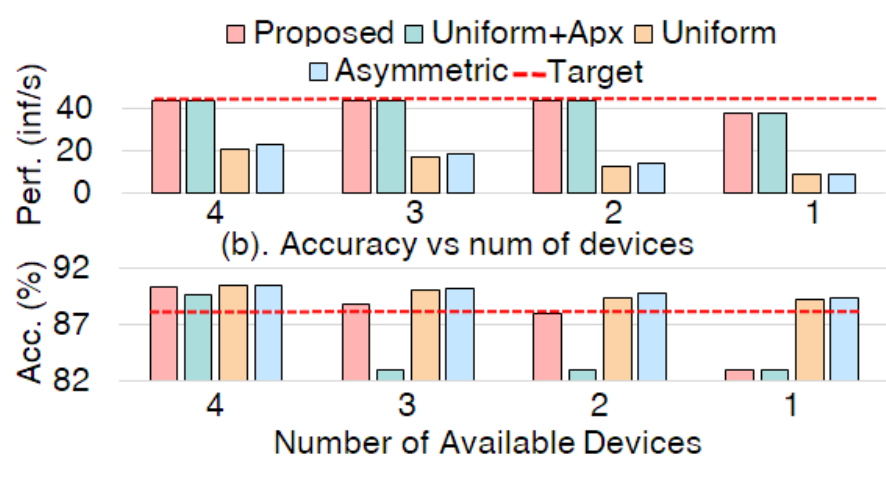}
\vspace{-9pt}
\caption{(a) Performance (inference per second) and (b) Accuracy (\%) vs. number of available devices.}
\label{fig.res_num_devices}
\vspace{-15pt}
\end{figure}

\noindent\textbf{Varying Device Availability.} Figure \ref{fig.res_num_devices} shows how each strategy copes with device unavailability/disconnection while meeting specific workload requirements for an input batch size of 650 images. We progressively disconnect one device from the cluster at run-time and present the output performance and accuracy in Figure \ref{fig.res_num_devices}
(a) and (b), respectively. The proposed approach meets both performance and accuracy requirements even with fewer available devices, by opportunistically exploiting accuracy tradeoffs. The \textit{Uniform} and \textit{Asymmetric} strategies fail to meet the performance requirements. The \textit{Uniform+apx} strategy meets the performance requirements with two devices; however, it fails to meet accuracy constraints with greedy approximation. 
On average, our approach is showing output performance and accuracy improvements by 63.98\% and 4.2\%.


\section{Conclusion}
We propose an adaptive workload distribution policy that partitions \ac{dnn} inference requests on collaborative heterogeneous edge clusters. Our approach exploits accuracy-performance trade-offs of \ac{dnn} models to jointly determine optimal partitioning points and accuracy levels of dynamic inference requests. We implemented our strategy on a real hardware testbed of a cluster of heterogeneous edge devices. We evaluated our proposed approach against other relevant workload distribution strategies and observed an average performance gain of 42.5\% and an average accuracy gain of 4.2\% against accuracy-aware solutions.

\bibliographystyle{IEEEtran}
\bibliography{references}

\end{document}